\documentclass[11pt]{amsart}

\usepackage{a4}
\usepackage[utf8]{inputenc}

\usepackage{amssymb}
\usepackage{hyperref}

\def\ci{{\mathcal I}}

\let\nonu=\nonumber

\def\nfrac#1#2{{\textstyle\frac{#1}{#2}}}

\begin{document}


\title{An Efficient Method for the Solution of Schwinger--Dyson equations for 
propagators. }
\author{Marc~P.~Bellon}
\address{UPMC Univ Paris 06, UMR 7589, LPTHE, F-75005, Paris, France}
\address{CNRS, UMR 7589, LPTHE, F-75005, Paris, France}
\date{}
\maketitle

\begin{abstract}
Efficient computation methods are devised for the perturbative solution of 
Schwinger--Dyson equations for propagators. I show how a simple computation 
allows to obtain the dominant contribution in the sum of many parts of previous 
computations. This allows for an easy study of the asymptotic behavior of the 
perturbative series. In the cases of the four-dimensional supersymmetric 
Wess--Zumino model and the $\phi_6^3$ complex scalar field, the singularities 
of the Borel transform for both positive and negative values of the parameter 
are obtained and compared.
\end{abstract}

\section{Introduction}\label{int}

In preceding works~\cite{BeSc08,Be10}, we have shown how to solve 
Schwinger--Dyson equations for propagators. This allows to compute the 
\(\beta\)--function in models without vertex divergencies, e.g., the 
supersymmetric Wess--Zumino model and some version of a six-dimensional theory 
of a scalar field with cubic interactions. The resolution method in the first 
of these works~\cite{BeSc08} is exact, but does not allow for an easy 
understanding of the properties of the computed perturbative series. This 
method is also computationally heavy and would scale badly for Schwinger--Dyson 
equations with a greater number of propagators.

The approximate differential equations introduced in~\cite{Be10} have the 
double advantage of allowing an easy derivations of the asymptotic behavior of 
the perturbative series and of a complexity which is only quadratic in the 
perturbative order. However, as can be seen in the second equation proposed for 
the Wess--Zumino model, this simplicity is lost if we have to get more precise 
results.

In this work, I propose a procedure which allows to systematically improve the 
second method, while remaining computationally simple and making the asymptotic 
behavior of the perturbative series transparent. In particular, I shall derive 
the dominant singularity of the Borel transform of the perturbative series on 
the positive axis, through the added precision on the asymptotic behavior of 
the perturbative series. In this work, I limit myself to the contributions of 
the dominant poles, which are sufficient for the asymptotic behavior, but the 
inclusion of additional terms would allow to reach any desired precision, for a 
fraction of the computational cost of the method used in~\cite{BeSc08}.

All these computations are based on two techniques, which have been described 
extensively in~\cite{BeSc08,Be10}, elaborating on the work of Kreimer and 
Yeats~\cite{KrYe2006}: a renormalization group equation allows to deduce the 
full propagator from the renormalization group functions and a Mellin transform 
of the one loop diagram gives the renormalization group functions from the 
propagators. These results will be freely used here.

In the present paper, we will see in a first step how the contribution of a 
simple pole in the Mellin transform of the diagram can be computed recursively, 
allowing the computation of an infinite sum of derivatives of the propagator 
while sidestepping the computation of the individual derivatives. This is 
immediately applied in the following section to the solution of the linear 
Schwinger--Dyson equation for a scalar model in dimension 6 which was first 
solved in~\cite{BrKr99}. The new method surely recovers the same numerical 
values and is computationally comparable, but it is far superior for the 
asymptotic analysis of the perturbative series and it prepares for the more 
complex examples. 

In the cases of non-linear Schwinger--Dyson equations, some terms cannot be 
computed with this first method and I extend it to the case where the 
denominator depends on the sum of some variables of the Mellin transform. This 
method is then applied to the supersymmetric Wess--Zumino model. The main 
result thus obtained is a precise asymptotic study of the perturbative series: 
beyond the dominant contribution which gives a pole for the Borel transform of 
the series, the exact nature of the singularity on the positive axis, which was 
discovered in the numerical computation of~\cite{BeSc08}, is uncovered. 
Finally, the scalar model in dimension 6 is studied in the case of a non-linear 
Schwinger--Dyson equation. Again, a singularity on the positive real axis is 
predicted, but it would have been difficult to spot numerically, since the non 
alternating part is smaller by a factor bigger than \(n^4\). Finally, I 
conclude on the possible extensions of this work, in particular with respect to 
the inclusion of higher loop corrections in the Schwinger--Dyson equations. The 
singularities of the Mellin transforms for higher loop diagrams are poles with 
denominators of the type envisioned here: the dominant contributions of the 
high order derivatives of the Mellin transform can be computed by the methods 
presented here.

\section{Contribution from a simple pole.} \label{simple}
We keep the fundamental insight in~\cite{Be10} that the Taylor series is 
dominated by the contribution of the singularities near the origin. In the 
one-loop case, the Mellin transform is expressible through \(\Gamma\) functions 
and these singularities are single poles. The first case to consider is 
therefore the one of a simple pole $1/(k+x)$.

The corresponding contribution to the $\gamma$-function is:
\begin{equation} \label{sum}
\frac 1 { k+x} = \frac 1 k \sum_{p=0}^\infty \left(\frac {-x}k\right)^p 
\longrightarrow 
	\ci = \frac 1k \sum_{p=0}^\infty  \left(\frac {-1}k\right)^p \gamma_p
\end{equation}
In this expression, $\gamma_p$ is the $p^{\mathrm th}$ derivative with respect 
to the logarithm of the impulsion $L$ of the correction to the propagator 
$G(L)$. However, the seemingly local sum of  derivatives in 
equation~(\ref{sum}) can be given an equivalent integral form:
\begin{equation}
	\ci_k = \int_{-\infty}^0 G(L) e^{kL}dL
\end{equation}
If $k$ is negative, the integral has to be taken from 0 to $+\infty$. For 
$k=1$, we recover the integral
\[ \int_0^{\mu^2} G(p^2) d(p^2), \]
which appeared in the work of Broadhurst and Kreimer~\cite{BrKr99}.

The point is that $\ci$ can be efficiently evaluated through the use of the 
renormalization group.
The action of $\gamma + \beta \partial_a$ on $G(L)$ gives its derivative with 
respect to $L$;
\begin{equation} \label{fund1}
(\gamma + \beta a  \partial_a) \ci_k =  \int _{-\infty}^0 \partial_L G(L) 
e^{kL} dL = G(0) - k \ci_k
\end{equation}
The last equality comes from an integration by part. Equation~(\ref{fund1}) can 
be used to obtain an efficient recursive computation of the perturbative 
expansion of $\ci$.

\section{The linear case.}
As a first application, let us go back to the linear Schwinger--Dyson equation 
for a 6-dimensional scalar field first studied in~\cite{BrKr99}. The 
one-variable Mellin transform is a rational function and a simple solution 
method was presented in~\cite{BrKr99}. This amounts to acting on the 
Schwinger--Dyson equation with an adequate differential operator in the 
logarithm of the momentum $L$ to reduce the right-hand side to a constant. 
In~\cite{Be10}, I proposed to act with a simpler differential operator. This 
allowed in particular to more easily account for the asymptotic behavior of the 
perturbative solution, but the truncated right-hand side precluded an exact 
solution. 

With the insight of the preceding section, it is possible to obtain the exact 
solution in a form which allows for an easy asymptotic analysis. Indeed, 
through the use of $\ci_2$ and $\ci_3$ which represent the effect of the 
propagator on the single poles $1/(2+x)$ and $1/(3+x)$, equation~(10) 
of~\cite{Be10} can be written:
\begin{equation}
	\gamma + \gamma(2a\partial_a -1) \gamma = a(\ci_2-\ci_3). \label{Lci}
\end{equation}
The recursive evaluation of $\ci_2$ and $\ci_3$ through the use of 
equation~(\ref{fund1}), completed by the identity $\beta=2\gamma$, allows for a 
rapid evaluation of the perturbative series for $\gamma$. The result coincides 
with the one stemming from the partial differential equation of~\cite{BrKr99}, 
providing a check of the computation.

It is then easy to obtain from equation~(\ref{Lci}) the asymptotic behavior of 
the perturbative series. Let us fix the notation for the perturbative 
coefficients of $\gamma$, $\ci_2$ and $\ci_3$:
\begin{equation}
 \gamma = \sum_{n=1}^\infty c_n a^n, \quad \ci_2 = \sum_{n=0}^\infty d_n a^n, 
\quad \ci_3 = \sum_{n=0}^\infty f_n a^n.
\end{equation}
It is convenient to convert the term bilinear in $\gamma$ to the form 
$(a\partial_a-1)\gamma^2$, so that with the first values $c_1=1/6$ and $c_2 = 
-11/6^3$, equation~(\ref{Lci}) gives the following relation:
\begin{equation}
 c_{n+1} \simeq -\frac {2n}{6} c_n + \frac {22(n-1)}{6^3} c_{n-1} + d_n - f_n.
\end{equation}
In the equation~(\ref{fund1}), the dominant components come from the cases 
where either $\gamma$ or $\ci_k$ is of maximal order, so that one obtains:
\begin{eqnarray}
d_n &\simeq& -\frac 14 c_n - \frac{2n-1}{12} d_{n-1} \\
f_n &\simeq& -\frac 19 c_n - \frac{2n-1}{18} f_{n-1} 
\end{eqnarray}
Now, it is consistent to suppose that $d_n$ and $f_n$ are proportional to $c_n$ 
and replace $c_{n-1}$ in the right hand side of the previous equations by the 
its approximate value $-\frac 3 n c_n$ to solve for the unknown factor. One 
obtains:
\begin{equation}
 d_n \simeq -\frac 1 2 c_n, \quad f_n \simeq -\frac 1 6 c_n.\label{ciasym}
\end{equation}
Putting all together, one obtains the following asymptotic relation:
\begin{equation}
 c_{n+1} \simeq- \left( \frac n 3 + \frac {23}{36}\right) c_n,
\end{equation}
which is well verified on the exact solution. By considering additional terms 
in the expansion of $\gamma^2$ and the recursive definition of $d_n$ and $f_n$, 
one could obtain additional terms in the expansion in $\frac 1n$ of the ratio 
of $c_n$ over $c_{n-1}$. 

\section{General poles.} \label{method}
When studying non-linear Schwinger--Dyson equations, Mellin transforms with 
many variables must be used. However, the obtained singularities are really 
simple, since they correspond to the divergences appearing when some subgraph 
becomes scale invariant: the singularities are simple poles whose denominators 
only depend on the sum of the Mellin variables associated to the subgraph.

The formulation of the pairing between propagator and Mellin transform appears 
quite dissymmetric, but it is in fact totally symmetric. The evaluation of a 
function of the propagator $G(L)$ can be reduced to the following pairing of 
$G(L)$ with the Mellin transform $H(x)$:
\begin{equation}
\label{pair} \sum_{n=0}^\infty \frac 1{n!} \partial_L^n G(L)\vert_{L=0} 
\partial_x^n H(x) \vert_{x=0}
\end{equation}
For simple enough Mellin transforms \(H\), it will be more convenient therefore 
to write this duality in reverse order, i.e., as some differential operator 
acting on the propagator \(G(L)\). The structure will simply be:
\begin{equation}
 H(\partial_{L}) G(L),
\end{equation}
since for an analytic \(H\), the Taylor series gives the value of the function. 
The situation becomes more interesting when there are many variables, since a 
sum of derivatives with respect to \(L_i\) can be converted to a simple 
derivative if we identify the different variables. The methods of 
section~\ref{simple} can then be applied to evaluate the effect of  a simple 
pole. 

However, one must pay special attention to the numerator: if in one variable, 
any numerator can be reduced to a scalar, it is not the case in a 
multidimensional setting. It is even possible, if we start from a multiloop 
diagram, that the numerator itself be a meromorphic function with poles. The 
identification of the different variables \(L_i\) has therefore to be done 
after the application of the differential operator corresponding to the part of 
the numerator which depends on the variables in the denominator. The final 
result is that we can evaluate easily the following type of pairings:
\begin{equation}
	\ci = \left.\frac1{k+\sum_{j=1}^n \partial_{L_j}} 
N(\partial_{L_1},\ldots,\partial_{L_n}) 
		\prod_{j=1}^n G(L_j)\right|_{L_1=\cdots=L_n=0}
\end{equation}
We obtain an equation similar to eq.~(\ref{fund1}), apart that the anomalous 
dimension \(\gamma\) is multiplied by the number of fields \(n\) and that 
\(G(0)\) is replaced by an expression involving \(N\).
\begin{equation}
\bigl( k + n \gamma + \beta a \partial_a\bigr) \ci =  \left. 
N(\partial_{L_1},\ldots,\partial_{L_n}) 
		\prod_{j=1}^n G(L_j)\right|_{L_1=\cdots=L_n=0}
\end{equation}

\section{Wess--Zumino model.}
Equipped with these new tools, let us consider the supersymmetric Wess--Zumino 
model. It is now possible to compute additional terms of the asymptotic 
behavior of the perturbative series. In particular, I shall obtain the exact 
nature of the singularity on the positive real axis. The aim is to include all 
the poles at unit distance from the origin,  that is to use the approximation 
of the Mellin transform introduced in equation~(21) of~\cite{BeSc08}, corrected 
to have the exact $xy$ term:
\begin{equation}
h(x,y) = (1+xy)\left(\frac1{1+x}+\frac1{1+y}-1\right) +
{\textstyle\frac12}\frac {xy}{1-x-y} + {\textstyle\frac12} xy. \label{asymp}
\end{equation}
To solve the Schwinger--Dyson equation with this approximate Mellin transform, 
we need two functions in addition to $\gamma$ itself: $F$, the sum of the 
$\gamma_n$ associated to $1/(1+x)$ and $H$, the function associated to the term 
$xy/(1-x-y)$. For the same approximation, the differential equation obtained in 
section~3.4 of~\cite{Be10} is very complex.

The renormalization group yields to the following functional equations for 
\(F\) and \(H\):
\begin{eqnarray}
\label{rgF} F &=& 1 - \gamma ( 3 a \partial_a +1 ) F,\\
\label{rgG} H &=& \gamma^2 + \gamma( 3 a \partial_a + 2) H.
\end{eqnarray}
Using the approximate Mellin transform~(\ref{asymp}), the Schwinger--Dyson 
equation then takes the form:
\begin{equation} \label{SDaWZ}
	\gamma = 2 a F - a - 2 a \gamma(F-1) + {\textstyle\frac12} a (H - 
\gamma^2)
\end{equation}
The three equations~(\ref{rgF},\ref{rgG},\ref{SDaWZ}) can be converted in 
recursion equations for the perturbative expansions of the functions $\gamma$, 
$F$ and~$G$.  

The numerical solution is easy. Compared with the complete calculations 
of~\cite{BeSc08}, the obtained precision is good, with errors which are always 
less then 1 percent and are asymptotically of the order of half a percent. This 
divides the error by a factor 10 with respect to the cruder approximation 
of~\cite{Be10} and moreover capture the non-alternating part in the development 
of $\gamma$: an asymptotic analysis of this system of equations will allow to 
obtain the exact nature of the singularity on the positive axis in the Borel 
transform of the series for $\gamma$.

Let us write $\gamma = \sum c_n a^n$, \(F = \sum f_n a^n\) and \(H= \sum h_n 
a^n\). It is easy to show that \(c_1=1\), \(c_2=-2\), \(f_2=1\). With the rapid 
growth of all the coefficients, one obtains the following asymptotic forms of 
the equations~(\ref{rgF},\ref{rgG}):
\begin{eqnarray}
\label{asymF} f_{n+1} &\simeq& -(3n+1) f_n +2(3n-2) f_{n-1}  - c_{n+1},\\
\label{asymG} h_{n+1} &\simeq& 2 c_n - 4 c_{n-1} + (3n+2) h_n - 2(3n-1)h_{n-1} 
+ 8 c_{n-1}.
\end{eqnarray}
In the first of these equations, $(3n-2)f_{n-1}$ and $c_{n+1}$ are respectively 
proportional to $-f_n$ and $2f_n$, yielding the asymptotic relation:
\begin{equation}
 f_{n+1} \simeq -(3n+5) f_n. \label{asy2F}
\end{equation}
Remarking that the dominant term in $c_n$ is $2f_{n-1}$, one recovers the 
dominant asymptotic relation for \(c_n\), \(c_{n+1} \simeq -(3n+2) c_n\), 
observed in~\cite{BeSc08} and proved in~\cite{Be10}. For $h_n$, one obtains 
similarly the asymptotic relation
\begin{equation}
h_{n+1} \simeq 3n h_n.
\end{equation}
This term is non-alternating and is the dominant contribution to the 
singularity of the Borel transform on the positive real axis. The ratio of the 
growth of \(f_n\) and \(g_n\) is in absolute value \(1-5/3n\). This translates 
in a ratio \(n^{-5/3}\) between the absolute values of these two contributions: 
this ratio coincides nicely with the numeric results and confirm the subleading 
character of the \(c_n\) term in eq.~(\ref{asymG}). 

\section{The \(\phi_6^3\) model.}

In this case,  one considers the action of \(\partial_L + \partial_L^2\) on the 
Schwinger--Dyson equation. The left-hand side is then
\begin{equation}
\gamma + \gamma(3 a \partial_a -1) \gamma,
\end{equation}
and the right-hand side is based on the following Mellin transform:
\begin{equation}
H(x,y) = \frac{ \Gamma(2+x) \Gamma(2+y) \Gamma(1-x-y)} { \Gamma(1-x) 
\Gamma(1-y) \Gamma(4+x+y)} 
\end{equation}
I approximate \(H\) with the contributions from the poles for \(x\) or \(y\) 
equal to \(-2\) or \(-3\) and the one for \(1-x-y=0\). For \(y=-2\), the 
residue is \((1-x)(1-x/2)\) and is converted to \((1+xy/2)(1+xy/4)\) not to mar 
the low order terms. For \(y=-3\), the residue has two more factors and is 
\(-(1+x)(1-x)(1-x/2)(1-x/3)\). For \(1-x-y=0\), the residue is 
\(xy/12(1+xy/2)\). We therefore will use the following approximation of \(H\):
\begin{eqnarray}
h(x,y) 
&=&(1+\frac{xy}2)(1+\frac{xy}4)\left(\frac1{2+x}+\frac1{2+y}-\frac12\right) 
\nonu \\
&& - 
(1-\frac{xy}3)(1+\frac{xy}3)(1+\frac{xy}6)(1+\frac{xy}9)\left(\frac1{3+x}+\frac1
{3+y}-\frac13\right) \nonu\\
&&+ \frac1{12} xy(1+\frac{xy}2) \frac1{1-x-y}- \frac 5{216}xy. \label{approx}
\end{eqnarray}
The last term in this equation adjust the coefficient of \(xy\) to the exact 
value, so that the cubic term in \(\gamma\) is exact. Most remarkably, this 
coefficient appears as the sum of the contributions from poles more distant 
from the origin, with poles in \(k+x\) giving negative contributions and the 
poles in \(k-x-y\) giving positive contributions. For the terms in \(x^2y\) or 
\(xy^2\) where both contributions add up, the remainder is larger. 

For the asymptotic analysis, these additional  terms do not matter. We need 
three functions in addition to the anomalous dimension \(\gamma\), \(\ci_2\) 
and \(\ci_3\) as in the case of the linear Schwinger--Dyson equation and a 
function \(H\) associated to the term \(xy(1+xy/2)/(1-x-y)\). We have similar 
equations for \(\ci_2\) and \(\ci_3\) than in the linear case:
\begin{eqnarray}
2 \ci_2 &=& 1 - \gamma( 1+ 3 a\partial_a) \ci_2, \label{ci2nl}\\
3 \ci_3 &=& 1 - \gamma( 1+ 3 a\partial_a) \ci_3. \label{ci3nl}
\end{eqnarray}
The recursion equation for \(H\) just has additional source terms with respect 
to the Wess--Zumino case, due to the more complex numerator:
\begin{equation}
H = \gamma^2 + \nfrac12 \gamma_2^2 + \gamma(2+3 a \partial_a) H.
\label{Hnl}
\end{equation}
Finally, using the approximate form of the Mellin transform~(\ref{approx}) and 
neglecting terms divisible by \( (xy)^3\) which are subdominant, one obtains 
the following equation:
\begin{eqnarray} \label{pnl}
\gamma+ \gamma(3a\partial_a-1)\gamma &=& 
a (2\ci_2-\nfrac1 2) + a\nfrac 3 4 \gamma (2\ci_2 -1 -\nfrac14 \gamma ) + 
a\nfrac 1 8 \gamma_2 ( 2\ci_2 -1 - \nfrac 1 2 \gamma) \nonu\\
&& - a (2\ci_3-\nfrac1 3) - \nfrac 5{18} a \gamma(2\ci_3-\nfrac2 
3-\nfrac19\gamma)  \nonu\\
&& + \nfrac 5{54} a \gamma_2 ( 2\ci_3 -\nfrac 2  3 -\nfrac 2 9 \gamma)
+ \nfrac 1{12} a H - \nfrac 5{216} a \gamma^2.
\end{eqnarray}
This gives the first two non-zero coefficients of \(\gamma\), \(c_1=\frac 1 
6\), \(c_2 =- \frac{11}{108}\). The dominant contribution for \(c_{n+1}\) comes 
from the second term of the left-hand side and is proportional to \(n/2\). The  
dominant terms for \(\ci_2\) and \(\ci_3\) are in fact exactly the same than in 
the case of the linear Schwinger--Dyson equation, since the changes from \(2n\) 
to \(3n\) in the recursions for \(c_n\) and  in 
equations~(\ref{ci2nl},\ref{ci3nl}) have compensating effects. 
Eq.~(\ref{ciasym}) remains valid at the dominant level. At the next to leading 
order, we have therefore:
\begin{equation}
\label{nonlinSD}
c_{n+1} \simeq -(3n+1) ( c_1 c_n + c_2 c_{n-1}) + 2 d_n - 2 f_n \simeq - (\frac 
n 2 + \frac {13} 9) c_n.
\end{equation}
The recursion operation for \(H\) (\ref{Hnl}) will differ from the Wess--Zumino 
case through the values of the first coefficients of \(\gamma\). The source 
terms are not important for the non-alternating component of \(H\). We obtain
\begin{equation}
 h_{n+1} \simeq \frac 16 (3n+2) h_n - \frac{11}{108}(3n-1) h_{n-1}\simeq (\frac 
n 2 - \frac 5{18})h_n . \label{hnl}
\end{equation}
The absolute values of the factors in the recursions for \(h_n\) and \(c_n\) 
are in the ratio \(1-31/(9n)\), meaning that the absolute value of \(h_n\) is 
smaller by a power \(31/9 = 3,444\dots\) of \(n\) than \(c_n\). Compounded with 
the fact that \(h_n\) intervenes only in \(c_{n+1}\), this non-alternating 
component in \(c_n\) could easily have been missed in a numerical study. In a 
sense, the situation is even worse, since the \(\gamma^2\) term in 
equation~(\ref{Hnl}) will give a larger contribution, so that \(h_n\) itself is 
dominated by this alternating component. However, this is not really a problem 
since the asymptotic recurrence relations are linear and the full solution will 
be a superposition of the particular solution proportional to \(c_{n-1}\) and a 
solution of the equation~(\ref{hnl}).

\section{Conclusion.}

In this work, the solution of Schwinger--Dyson equations for propagators has 
made a new step forward: I reach not only the leading asymptotic behavior of 
the perturbative terms, but the subleading, ``wrong sign" contribution, as well 
as systematic corrections in powers of \(1/n\). These computations can be made 
easily more precise through the inclusion of additional poles of the Mellin 
transform or the adjunction of a few monomials: a suitable combination of the 
two methods should allow to obtain the solution of the Schwinger--Dyson 
equations with very high precision for a fraction of the computational cost of 
the methods of~\cite{BeSc08}.  However, such a full precision computation is 
not really useful since the Schwinger--Dyson equations I consider here are but 
the first approximation to the full system of Schwinger--Dyson equations. 

In the computation of higher order corrections to the Schwinger--Dyson 
equations, the power of the present methods should be precious. Indeed, in a 
higher loop primitively divergent diagram, the number of individual propagators 
is higher and the full evaluation  through the straightforward methods 
of~\cite{BeSc08}, much more complex. For a Mellin transform in \(k\) variables, 
the number of different derivatives up to total order \(n\) scales as 
\(n^{k-1}\), with the evaluation of each of these terms at perturbative order 
\(2n\) again implying approximatively \(n^{k-1}\) operations. Some time for 
space bartering could reduce somewhat this growth by reusing partial products, 
but in any case, this complexity would make any computation, whether explicit 
or asymptotic, very complex, even if the evaluation of the derivatives were not 
difficult per se. 

In the approach presented here, the situation is much more manageable: the pole 
structure of the Mellin transform, which is linked to the divergences of 
subdiagrams, is of the form analyzed in section~\ref{method} and their 
contributions can be computed in linear time. Furthermore, one can identify the 
residues which are important for the asymptotic behavior of the perturbative 
series and focus the analytical evaluations on them. This will be the subject 
of a forthcoming publication~\cite{Be10b}. If, up to now, these works have 
dealt only with Schwinger--Dyson equations for propagators, the efficiency of 
the methods introduced here should be important in the more challenging cases 
of systems including Schwinger--Dyson equations for vertices.

In the spirit of~\cite{Be10}, we could also deduce systems of differential 
equations for the anomalous dimension and some auxiliary functions, with the 
hope of obtaining information on the asymptotic behavior of the anomalous 
dimension at large coupling. However, the singularity of the Borel transform on 
the positive axis indicates that the Borel resummation is not uniquely defined. 
It is therefore a challenge to determine whether such a system of differential 
equations, which would be determined from purely perturbative considerations, 
can be given a non-perturbative meaning.


\begin{thebibliography}{1}

\bibitem{BeSc08}
{M}arc {B}ellon and {F}idel {S}chaposnik.
\newblock Renormalization group functions for the {W}ess-{Z}umino model: up to
  200 loops through {H}opf algebras.
\newblock {\em Nucl.\ Phys.\ B}, 800:517--526, 2008.
\newblock \href {http://arxiv.org/abs/arXiv:0801.0727v2 [hep-th]}
  {\path{arXiv:arXiv:0801.0727v2 [hep-th]}}.

\bibitem{Be10}
{M}arc~P. {B}ellon.
\newblock {A}pproximate {D}ifferential {E}quations for {R}enormalization
  {G}roup {F}unctions in {M}odels {F}ree of {V}ertex {D}ivergencies.
\newblock {\em Nucl.\ Phys.\ B}, 826 [{PM}]:522--531, 2010.
\newblock \href {http://arxiv.org/abs/0907.2296} {\path{arXiv:0907.2296}},
  \href {http://dx.doi.org/10.1016/j.nuclphysb.2009.11.002}
  {\path{doi:10.1016/j.nuclphysb.2009.11.002}}.

\bibitem{Be10b}
{M}arc~P. {B}ellon.
\newblock Higher loop corrections to {S}chwinger-{D}yson equations.
\newblock In preparation, 2010.

\bibitem{BrKr99}
D.~J. Broadhurst and D.~Kreimer.
\newblock Exact solutions of {D}yson--{S}chwinger equations for iterated
  one-loop integrals and propagator-coupling duality.
\newblock {\em Nucl.\ Phys.}, B 600:403--422, 2001.
\newblock \href {http://arxiv.org/abs/hep-th/0012146}
  {\path{arXiv:hep-th/0012146}}.

\bibitem{KrYe2006}
Dirk Kreimer and Karen Yeats.
\newblock {An etude in non-linear Dyson-Schwinger equations}.
\newblock {\em Nucl. Phys. Proc. Suppl.}, 160:116--121, 2006.
\newblock \href {http://arxiv.org/abs/hep-th/0605096}
  {\path{arXiv:hep-th/0605096}}, \href
  {http://dx.doi.org/10.1016/j.nuclphysbps.2006.09.036}
  {\path{doi:10.1016/j.nuclphysbps.2006.09.036}}.

\end{thebibliography}
\end{document}